# Testbed for Wireless Vehicle Communication: a Simulation Approach based on Three-Phase Traffic Theory

B. S. Kerner, S. L. Klenov and A. Brakemeier

*Abstract*— A testbed for wireless vehicle communication based on a microscopic model in the framework of three-phase traffic theory is presented. In this testbed, vehicle motion in traffic flow and analyses of a vehicle communication channel access based on IEEE 802.11e mechanisms, radio propagation modeling, message reception characteristics as well as all other effects associated with ad-hoc networks are integrated into a three-phase traffic flow model. Thus simulations of both vehicle ad-hoc network and traffic flow are integrated onto a single testbed and perform simultaneously. This allows us to make simulations of ad-hoc network performance as well as diverse scenarios of the effect of wireless vehicle communications on traffic flow during simulation times, which can be comparable with real characteristic times in traffic flow. In addition, the testbed allows us to simulate cooperative vehicle motion together with various traffic phenomena, like traffic breakdown at bottlenecks. Based on simulations of this testbed, some statistical features of ad-hoc vehicle networks as well as the effect of C2C communication on increase in the efficiency and safety of traffic are studied.[1]

## I. INTRODUCTION

Wireless vehicle communication, which is the basic technology for ad-hoc vehicle networks, is one of the most important scientific fields of future ITS. This is because there are many possible applications of ad-hoc vehicle networks, including various systems for danger warning, traffic adaptive assistance systems, traffic information and prediction in vehicles, improving of traffic flow characteristics through adaptive traffic control, etc. [1-7].

However, the evaluation of ad-hoc vehicle networks requires many communicating vehicles moving in real traffic flow, i.e., field studies of ad-hoc vehicle networks are very complex and expensive. For this reason, to prove the performance of ad-hoc vehicle networks based on wireless vehicle communication, reliable simulations of ad-hoc vehicle networks are of great importance and indispensable.

An usual schema for the development of a testbed for simulations of ad-hoc vehicle networks includes a traffic flow model, a model for vehicle communications that is often based on the use of ns-2 simulator [8], and application models (applications in Fig. 1) (e.g., [9-16]). Application models determine, for example, necessary changes in vehicle behavior in traffic flow after receiving of the associated message or/and whether this message should be resent to other vehicles or not. There are two different networks in such testbeds: (i) a traffic network simulated with the use of the traffic flow model and (ii) a communication (ad-hoc) network simulated with the use of the communication model in which positions and other characteristics of each communicated vehicle are taken from simulations of the traffic network made at the latest point in time. Simulations of many communicating vehicles in the communication network with known communication models are very time intensive. For this reason, often the model of communication network (communication model in Fig. 1) performs simulations based on traffic flow data previously simulated through the use of the traffic flow model (off-line simulations of traffic networks). In some of these testbeds, to study applications in which vehicle behavior should be changed in accordance with received messages, the communication model performs simulations after each time step of traffic flow simulations. In any case, the use of this simulation schema (Fig. 1) requires a very long run time of the simulations, which can be some order of magnitude longer than real time of vehicle moving in traffic flow.

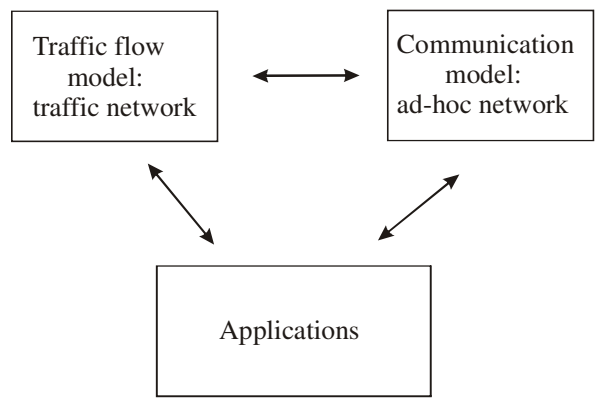

Fig. 1. An usual schema of testbeds for simulations of ad-hoc vehicle networks (e.g., [9-16]).

In this article, we suggest an approach for testbed development in which simulations of a traffic network and an ad-hoc vehicle network as well as applications are integrated into an united network, i.e., there is only one network in this testbed. The network describes both vehicle motion in traffic flow and communications as well as the effect of applications on traffic flow and vehicle behavior. As a result, simulations of ad-hoc performance and various applications can be made many times quicker than with the schema shown in Fig. 1. To reach this goal, each vehicle in this network exhibit different *attributes* needed for both vehicle motion and communications, and application scenarios. In addition, we should note that recently based on a study of measured data on many highways in different countries a three-phase traffic theory has been developed. In contrast with earlier traffic flow theories and models, three-phase traffic theory can explain and predict all known empirical features of traffic breakdown and resulting congested patterns [17]. For this reason, in the testbed presented in this article, we use a traffic flow model in the framework of three-phase traffic theory.

[1] B. S. Kerner and A. Brakemeier are with Daimler Research, HPC: 050 – G021, D-71059 Sindelfingen, Germany, e-mails: (boris.kerner, a.brakemeier) @daimler.com; S.L. Klenov is with Moscow Institute of Physics and Technology, Department of Physics, 141700 Dolgoprudny, Moscow Region, Russia (sergey_klenov@inbox.ru).

A testbed model is presented in Sect. II. Simulations of two scenarios of C2C application devoted to ad-hoc network performance and to a study of the influence of C2C communication on traffic flow are presented in Sects. III and IV, respectively.

## II. United Network Model for Simulations of C2C-Communication, Ad-Hoc Vehicle Networks and Traffic Flow

In the united network model of traffic flow, C2C-comunications, and ad-hoc networks, there are vehicle attributes, which exhibit each communicated vehicle (Fig. 2). All other vehicles in the network, which cannot communicate, exhibit only one attribute: update rules for vehicle motion. If in addition with communicated vehicles the network includes roadside communication units (RSU), each RSU exhibits the communicated vehicle attributes with the exclusion of the update rules for vehicle motion.

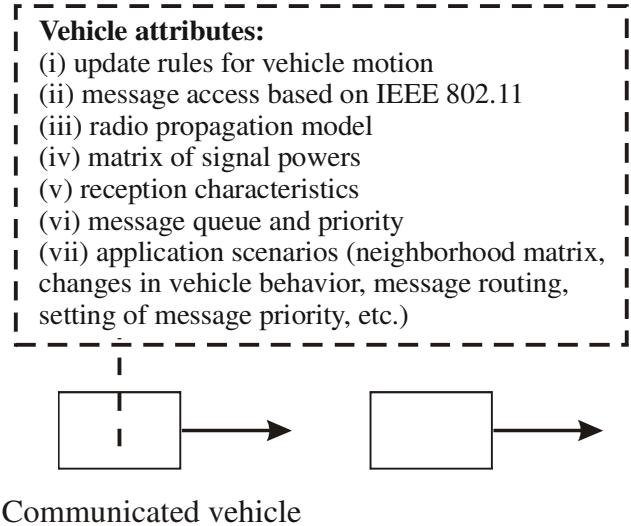

**Vehicle attributes:**
(i) update rules for vehicle motion
(ii) message access based on IEEE 802.11
(iii) radio propagation model
(iv) matrix of signal powers
(v) reception characteristics
(vi) message queue and priority
(vii) application scenarios (neighborhood matrix, changes in vehicle behavior, message routing, setting of message priority, etc.)

Communicated vehicle

Fig. 2. A schema of the testbed for simulations of ad-hoc networks and traffic flow within a united network model as presented in the article.

### 2.1. Update Rules for Vehicle Motion

The vehicle attribute "update rules for vehicle motion" are given by a stochastic microscopic three-phase traffic flow model of Kerner and Klenov [18, 19]. We discuss only a hypothetical traffic flow with identical vehicles. The Kerner-Klenov model of heterogeneous flow can be found in [17].

Basic rules of vehicle motion in the model are as follows:

$$v_{n+1} = \max(0, \min(v_{free}, \tilde{v}_{n+1} + \xi_n, v_n + a\tau, v_{s,n})),$$
$$x_{n+1} = x_n + v_{n+1}\tau, \tag{1}$$

where

$$\tilde{v}_{n+1} = \max(0, \min(v_{free}, v_{s,n}, v_{c,n})), \tag{2}$$

$$v_{c,n} = \begin{cases} v_n + \Delta_n & \text{for } g_n \leq G_n \\ v_n + a_n\tau & \text{for } g_n > G_n \end{cases}, \tag{3}$$

$$\Delta_n = \max(-b_n\tau, \min(a_n\tau, v_{\ell,n} - v_n)), \tag{4}$$

index $n$ corresponds to the discrete time $t = n\tau$, $n = 0,1,2,...$; $\tau$ is the time step; $v_{free}$ is the maximum speed in free flow; $\tilde{v}_{n+1}$ is the vehicle speed without noise component $\xi_n$ (see below); $g_n = x_{\ell,n} - x_n - d$ is the space gap (net distance) between vehicles, $d$ is the vehicle length; the lower index $\ell$ marks functions (or values) related to the preceding vehicle; $a_n \geq 0$ and $b_n \geq 0$ (see below); $G_n$ is a synchronized gap:

$$G_n = G(v_n, v_{\ell,n}), \tag{5}$$

$$G(u, w) = \max(0, k\tau u + \phi a^{-1} u(u - w)), \tag{6}$$

where $k>1$ and $\phi$ are constants.

In (1), $v_{s,n} = \min(v_n^{(safe)}, g_n/\tau + v_\ell^{(a)})$ is a safe speed, where $v_n^{(safe)}$ is a safe speed of the Krauß-model [20] that is a solution of the Gipps-equation [21]

$$v_n^{(safe)}\tau + X_d(v_n^{(safe)}) = g_n + X_d(v_{\ell,n}), \tag{7}$$

$X_d(u) = b\tau^2(\alpha\beta + \alpha(\alpha-1)/2)$ [20], $b$ is constant, $\alpha$ is the integer part of $u/b\tau$, $\beta$ is the fractional part of $u/b\tau$; $v_\ell^{(a)}$ is an anticipation speed (formula (16.48) of [17]).

To simulate driver time delays either in vehicle acceleration or in vehicle deceleration, $a_n$ and $b_n$ in (3), (4) are taken as the following stochastic functions

$$a_n = a\theta(P_0 - r_1), \tag{8}$$

$$b_n = a\theta(P_1 - r_1), \tag{9}$$

$$P_0 = \begin{cases} p_0 & \text{if } S_n \neq 1 \\ 1 & \text{if } S_n = 1, \end{cases} \tag{10}$$

$$P_1 = \begin{cases} p_1 & \text{if } S_n \neq -1 \\ p_2 & \text{if } S_n = -1, \end{cases} \tag{11}$$

where a is the maximum acceleration; $r_1 = \text{rand}(0,1)$, i.e., this is an independent random value uniformly distributed between 0 and 1; $\theta(z) = 0$ at $z < 0$ and $\theta(z) = 1$ at $z \geq 0$; probabilities $p_0(v)$, $p_2(v)$ are given functions of speed, probability $p_1$ is a model parameter; $1-P_0$ and $1-P_1$ are the probabilities of a random time delay in vehicle acceleration and deceleration, respectively; $S_n$ denotes the state of vehicle motion ($S_n = -1$ represents deceleration, $S_n = 1$ acceleration, and $S_n = 0$ motion at nearly constant speed):

$$S_n = \begin{cases} -1 & \text{if } \tilde{v}_n < v_{n-1} - \delta \\ 1 & \text{if } \tilde{v}_n > v_{n-1} + \delta \\ 0 & \text{otherwise,} \end{cases} \quad (12)$$

where $\delta$ is constant ($\delta \ll a\tau$).

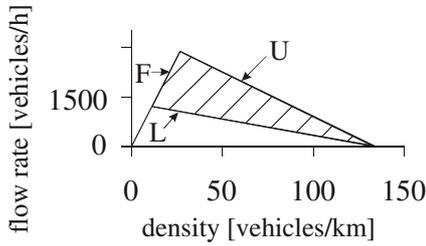

Fig. 3. Hypothetical steady model states [17-19].

Eqs. (2)-(6) describe the speed adaptation effect in synchronized flow. The speed adaptation effect within the synchronization gap means that hypothetical steady model states of synchronized flow (in which all vehicles move at the same time-independent speed and at the same space gap to one another) cover a 2D (two-dimensional) region in the flow-density plane (Fig. 3), i.e., there is no fundamental diagram for steady speed states of synchronized flow. The boundaries of this 2D region F, L, and U are respectively associated with free flow (F), the synchronization gap (L), and a safe gap (U) determined through the safe speed.

The noise component $\xi_n$ in (1) that simulates random deceleration and acceleration is applied depending on whether the vehicle decelerates or accelerates, or else maintains its speed:

$$\xi_n = \begin{cases} -\xi_b & \text{if } S_{n+1} = -1 \\ \xi_a & \text{if } S_{n+1} = 1 \\ 0 & \text{if } S_{n+1} = 0, \end{cases} \quad (13)$$

where $\xi_a$ and $\xi_b$ are random sources for deceleration and acceleration, respectively:

$$\xi_a = a\tau\theta(p_a - r), \quad (14)$$
$$\xi_b = a\tau\theta(p_b - r), \quad (15)$$

$p_a$ and $p_b$ are probabilities of random acceleration and deceleration, respectively; $r = \text{rand}(0,1)$.

In all simulations presented below we used a model of two-lane freeway sections with two directional traffic flows. Lane changing rules and bottleneck models can be found in [17]. Open boundary conditions are used for both directions. The flow rate of vehicles entering roads at the beginning of each of the directions $q_{in}$ is a parameter of simulations.

### 2.2. Message Access

During a motion of a communicating vehicle in the network calculated through the use of the update rules for vehicle motion (1)-(15), the vehicle (and RSU) attribute "message access", which is based on IEEE 802.11 basic access method [22-24], calculates vehicle access possibility for message sending at each time instant. At the end of the backoff procedure of the IEEE 802.11 access method (Fig. 4), a decision whether the medium is free or busy is made based on a vehicle (or RSU) attribute "matrix of signal powers" (see below). If there are messages to be send and the medium is free, the vehicle sends the message that has the highest priority and/or is the first one in the message queue in this vehicle.

### 2.3. Radio Propagation Model

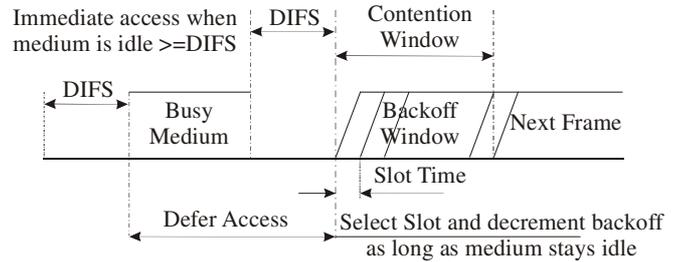

Fig. 4. IEEE 802.11 basic access mechanism [10, 22-24].

Based on the vehicle (and RSU) attribute "radio propagation model", signal powers of the message that has been sent by the vehicle are calculated for current locations of all other communicating vehicles and RSUs. If the related signal power is greater than a given threshold signal power $P_{th}$ (model parameter), then this signal power of the associated message is stored into a "matrix of signal powers" of the receiving vehicles. In all simulations presented, we have used a well-known two-ray-ground radio propagation model with communication range 200 m [25].

### 2.4. Matrix of Signal Powers

As mentioned, the event "the vehicle (or RSU) has sent the message" is registered and stored (together with the time instant when the associated signal power has been valid) in matrixes of signal powers of all other communicating vehicles and RSUs for which the associated signal power is greater than the threshold denoted by $P_{th}$, which is a model parameter.

This threshold $P_{th}$ is much smaller than the threshold CSTh (carrier sense threshold) of the IEEE 801.11 mechanism. The smaller $P_{th}$ is chosen, the greater the accuracy of simulations of ad-hoc network performance, however, the longer the simulations run time. In all simulation results presented below we have used $P_{th} = -116$dBm, which allows us to have a good balance between accuracy and the simulation time (simulation results are changed in the range of about 1%, when instead of the threshold $P_{th} = -116$dBm, the threshold $P_{th} = -126$dBm has been used).

### 2.5. Reception Characteristics

Signal reception characteristics are associated with an analysis of the vehicle (and RSU) attribute "matrix of signal powers", which is automatically made at each time instant for each communicating vehicle (or RSU) individually.

In particular, the vehicle (and RSU) attribute "matrix of signal powers" is used for the decision whether the medium is free or busy at each time instant as well as for the decision whether the vehicle (or RSU) has received a message or not. At each time instant, the matrix consists of signal powers, which are greater than the threshold $P_{th}$ associated with all other communicating vehicles and RSU in the network. If the sum of these powers is greater than the carrier sense threshold CSTh, the medium is considered to be busy; otherwise the medium is free.

To decide whether the vehicle (or RSU) has received a message or not at a given time instance, the highest power within the matrix of signal powers is considered. If this signal power is smaller than a receiving threshold RXTh, then the message is not accepted. If this signal power is greater than RXTh, then it is tested for the matrix of the signal power whether the ratio between the power of this signal and the sum of the powers of all other signals stored in the matrix is greater than the required signal-to-noise ratio (SNR) at the selected data rate (DR) for the whole duration of the message; If yes, the signal could be considered to be received, otherwise there is no message acceptance at the time instance.

We see that at each time instant the matrix of signal powers is used both for the decision whether the vehicle (or RSU) has received a message and whether there are collisions between two or more different signals at the current vehicle location (or RSU location). Message collisions are realized for example, if there are two or more signals within the matrix and the highest power is greater than the threshold RXTh, however, based on the above procedure the decision has been made that there is no message acceptance at the time instance. The decision about signal collisions is further used for a study of ad-hoc network performance.

### 2.6. Message Queue and Priority

Based on an application, which should be simulated, in the model each communicating vehicle (or RSU) exhibits an attribute of message queue organization and individual message priority performance governed automatically. Because each communicating vehicle or RSU manages these features individually, this attribute can be chosen differently for various types of the communicating vehicles or RSUs.

### 2.7. Application Scenarios

In the model, each communicating vehicle (and RSU) exhibits an attribute "application scenario". This attribute governs the organization of all messages that received and to be sent. Based on this attribute and the message context just received by the vehicle, the vehicle can change its behavior in traffic flow (e.g., the vehicle slows down or changes the lane, or else changes the route, etc.).

### III. APPLICATION SCENARIO 1: NEIGHBOR TABLE OF COMMUNICATING VEHICLES

In many applications of C2C-communication, an additional vehicle attribute is required: a matrix of neighborhood for communicating vehicles, also called neighbor table. In this matrix, current locations together with ID (and other information) of all communicating vehicles are stored, which are in direct communication range. One of the aims of the neighbor table is to perform the message routing in the ad-hoc network.

In simulations made we have assumed that with the aim of creating the neighbor table, each communicating vehicle tries to transmit a high priority message (beacon), which is created within the vehicle with a periodicity of 0.1 sec. Each communicating vehicle has in addition 49 other non-priority messages within the message queue. These non-priority messages should also be sent in intervals of 0.1 s. When, before a new non-priority message should be sent, a new beacon message is generated in the vehicle, the beacon message is put at the first place within the message queue, i.e., instead of the non-priority message, the beacon message applies for channel access.

The neighbor table should include all communicating vehicles in communication range. But due to signal collisions not all beacons are received and the corresponding vehicles do not appear in the neighbor table. We can see (Fig. 5) that the probability of these mistakes in the neighbor tables (right, curves 1) increases rapidly with the percentage of communicating vehicles $\eta$, whereas the probability of the message receiving (right, curves 2) only slightly decreases with $\eta$. We should note that at any $\eta$ under consideration the probability of mistakes in the matrix is not equal to zero (Fig. 5, right, curves 1).

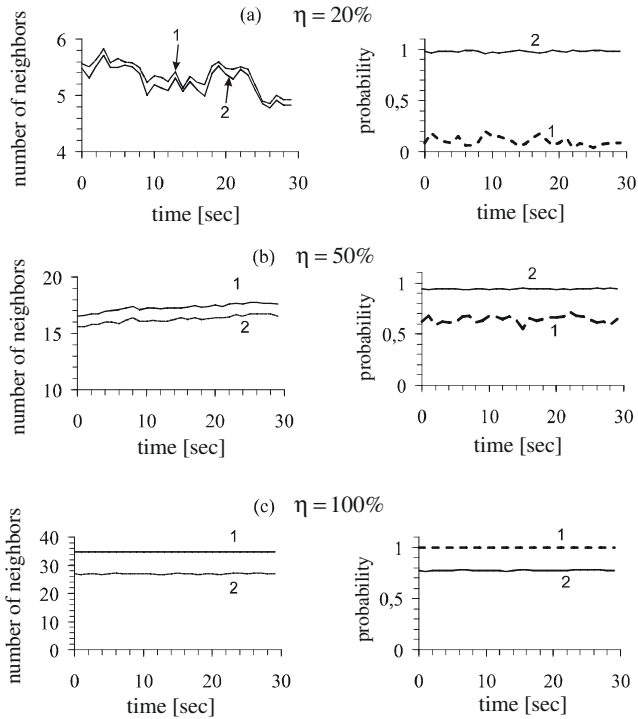

Fig.5. Performance characteristics of the neighbor table for different percentages of communicating vehicles $\eta$. Figures left: real number of communicating neighbors for a vehicle (curves 1) and the average number of communicating neighbors in the neighbor table (curves 2). Figures right: probabilities for one or more mistakes in the neighbor tables (curves 1) and for message receiving from one of the neighbors (curves 2). Contention window and AIFSD ([23]) for priority and non-priority messages, respectively: 7 and 15 slots, 45 and 58 $\mu$s. Slot is 13 $\mu$s. $CSTh = -96$dBm. $RXTh = -90$dBm. $SNR=6$dB. $DR=3$ Mb/s. Message length is 500B. $q_{in} = 2000$ vehicles/h/lane. Simulation time 30sec.

IV. APPLICATION SCENARIO 2: INFLUENCE OF C2C-COMMUNICATION ON CONGESTED TRAFFIC PATTERNS

To study the ad-hoc network influence on traffic flow characteristics, we consider the following scenario. There are two-lane freeway sections each for one traffic direction. In one of the direction, there is an on-ramp bottleneck at location 16 km. The flow rate on the main road upstream of the bottleneck is $q_{in} = 1946$ vehicles/h/lane; the flow rate to the on-ramp is $q_{on} = 300$ vehicles/h. At these flow and bottleneck parameters and if there is *no* C2C communication, a general congested pattern (GP) [17] occurs at the bottleneck. Features of GP formation are as follows: (i) firstly traffic breakdown, i.e., a phase transition from free flow to synchronized flow ($F \rightarrow S$ transition) occurs at the bottleneck (Fig. 6 (a)); (ii) the synchronized flow propagates upstream; (iii) within the flow, a pinch region of great density is formed within which narrow moving jams emerge and propagate upstream growing in their amplitude (Fig. 6 (b)); (iv) finally, the jams transform into wide moving jams (Fig. 6 (c)). Thus the GP consists of synchronized flow and wide moving jams that emerge within the synchronized flow (Fig. 6 (c, d)).

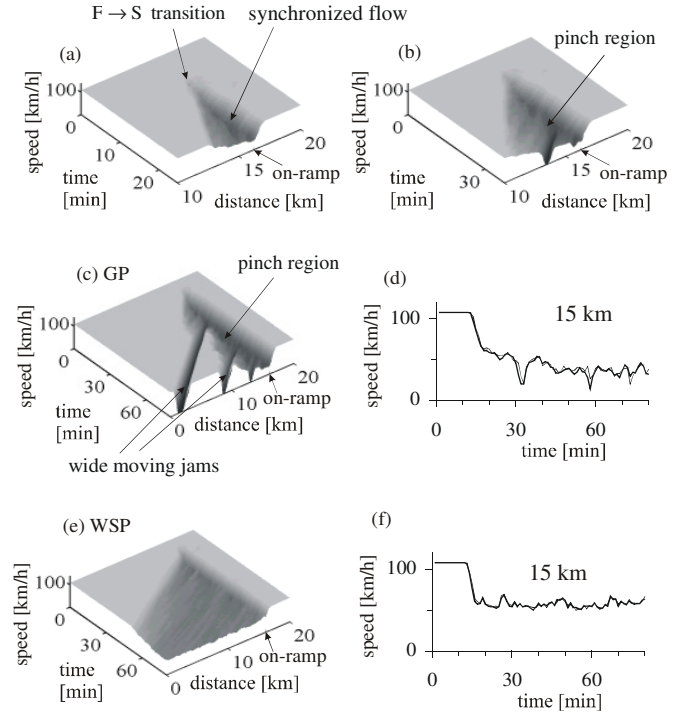

Fig.6. Simulation of influence of C2C communication on congested patterns at on-ramp bottleneck: (a-c, e) speed in time and space; (d, f) time functions of speed for GP (d) and WSP (f) at location 1 km upstream of the bottleneck.

Now we assume that all vehicles are communicated vehicles, which try to send 49 non-priority messages with time intervals 0.1 sec. When the $F \rightarrow S$ transition is occurring at the bottleneck (Fig. 6 (a)), then vehicles that pass the bottleneck generate and send a priority message about required change in the driver behavior within the synchronized flow. All vehicles that received the message resend it as the priority one. Simulations show that at chosen flow rates, there are almost no message collisions and therefore vehicles upstream of the bottleneck receive the priority message during several seconds with probability that is approximately equal to one.

We suggest that the message comprises a required space gap, which should be maintained by vehicles moving within the synchronized flow. This required space gap is generated in accordance with synchronized flow speed measured by vehicles passing the bottleneck.

In vehicle motion rules of the model, the associated change in driver behavior is simulated through an increase in probability $p_1$ in (11) from $p_1=0.3$ for vehicles, which have no information about the required space gap to $p_1=0.55$ for the vehicles that received the message. The greater $p_1$, the greater the difference between vehicle space gap and a safe space gap associated with the upper boundary in the 2D-region of synchronized flow states (boundary U in Fig. 3); in turn, the greater this gap difference, the less the probability for moving jam emergence in the synchronized flow [17].

As a result of space gap increase within the synchronized flow, the pinch region and therefore GP do not occur. Instead, a widening synchronized flow pattern (WSP) is forming (Fig. 6 (e, f)). Whereas in the pinch region of the GP the mean space gap is 15 m, it is 25 m within the WSP. Due to the transformation of the GP into the WSP, two effects are achieved: (i) wide moving jams do not occur and (ii) the average speed within synchronized flow upstream of the bottleneck increases from about 40 km/h within the GP to 60 km/h within the WSP. These effects result in a considerable increase in the efficiency and safety of traffic.

## V. DISCUSSION

1. Simulations made with the use of the testbed for ad-hoc networks presented in this paper allow us to perform quick simulations of various applications of C2C-communication and ad-hoc network performance associated with the real behavior of vehicular traffic. This is due of the following advantages of this testbed:

(i) As in a real ad-hoc network, there is only one network in the testbed in which C2C-communication, ad-hoc performance, and traffic flow characteristics are simulated simultaneously during vehicle motion. This testbed feature decreases the simulation run time considerably and exhibits a sufficient accuracy of simulations.

(ii) Moreover, this testbed feature allows us to make an easier understanding of ad-hoc network and traffic flow performances associated with those applications in which message contexts should influence on vehicle behavior. This is crucial especially for communication based safety systems that currently are studied in various research projects (e.g. WILLWARN [5] and SAFESPOT [26])

(iii) Vehicle motion is simulated based on a stochastic three-phase traffic flow model, which, as shown in the book [17], explains and predicts all known empirical (measured) spatiotemporal features of traffic.

2. Simulations show that C2C communication can increase the efficiency and safety of traffic considerably.

## ACKNOWLEDGEMENTS

We thank Gerhard Nöcker, Andreas Hiller and Christian Weiss for fruitful discussions.